\documentstyle[12pt]{article}
\begin{document}
\baselineskip 20pt 
\noindent
\hspace*{12.3cm}
{\bf (hep-ph/9908435)}\\
\noindent
\hspace*{13.5cm}
YUMS 99-021\\  
\noindent
\vspace{1.8cm}

\begin{center}
{\Large \bf Hierarchical Quark Mixing and Bimaximal Lepton Mixing 
on the Same Footing } 

\vspace*{1cm}

{\bf C. S. Kim \footnote{kim@cskim.yonsei.ac.kr,~~
http://phya.yonsei.ac.kr/\~{}cskim}}~~~~{\bf and}~~~~
{\bf J. D. Kim \footnote{jdkim@theory.yonsei.ac.kr}}

\vspace*{0.5cm}

 Physics Department, Yonsei University, Seoul 120-749, Korea \\ 
        
\vspace{1.0cm}
 
(\today)
\vspace{0.5cm}

\end{center}     

\begin{abstract}
\vspace{0.2cm}

\noindent
We show that  not only the hierarchical
quark CKM mixing matrix but also the ``bimaximal" lepton flavor 
mixing matrix can be derived
from the same mass matrix ansatz based on the broken permutation 
symmetry, by assuming
the hierarchy of neutrino masses to be $m_1\simeq m_2 <<m_3$.
We also reproduce the recently measured angle of unitary triangle,
$\sin 2\beta$, as well as all the observed experimental values of
$V_{\mbox{\tiny CKM}}$ of the quark CKM matrix.
And we predict Jarlskog rephasing invariant quantity,
$J_{\mbox{\tiny CP}} \simeq 0.18 \times 10^{-4}$, 
and the upper bound of the same quantity in the lepton sector, 
$J^l_{\mbox{\tiny CP}} \leq 0.012$, which may be indeed
large enough to generate the lepton number violation of the universe.

\end{abstract}

\vspace{1.0cm}
PACS number(s): 11.30Er,12.15Ff,14.60Pq

\newpage
\section{Introduction}

The flavor mixing and fermion masses and their hierarchical 
patterns remain to be one of the basic problems in particle physics.
Within the Standard Model, all masses and flavor mixing angles  
are free parameters and no relations among them are provided.
As an attempt to derive relationship between the quark masses and quark 
mixing hierarchies,
a quark mass-matrix {\it Ansatz} was suggested about two decades 
ago \cite{wein}.
This in fact reflects the  calculability \cite{cal}
of the flavor mixing angles in terms of the quark masses.
Of several {\it Ansatz} proposed, the canonical mass matrices of the 
Fritzsch type \cite{wein,test} and its variations \cite{cuypers}
had been generally assumed to predict the 
hierarchical Cabibbo-Kobayashi-Maskawa (CKM) matrix \cite{km} 
or the Wolfenstein mixing matrix \cite{wolf} except the unexpectedly
heavy top-quark mass, $m_t$.

For the lepton sector, we now have some evidence for lepton flavor mixing.
The observations of the solar neutrino deficit \cite{solar1,solar2}
and the atmospheric neutrino deficit \cite{superk1}
can be explained by neutrino oscillations, and
these in turn indicate nonzero neutrino masses and mixing.
The elements of the lepton flavor mixing matrix are determined from
the neutrino oscillation experiments.
The results of the atmospheric neutrino experiments can be explained 
by maximal mixing
between the muon neutrino and tau neutrino states, 
and those of the solar neutrino
experiments can be explained by vacuum oscillation with maximal mixing
between the electron neutrino and muon neutrino, 
although they may also be
explained through matter enhanced neutrino oscillation.
These results imply the ``bimaximal" mixing pattern between three flavor 
neutrinos \cite{bimax}.

Thus, it is likely that the ``bimaximal" mixing pattern of 
the lepton sector is 
quite different from the hierarchical mixing of the quark sector. 
At the first glance, the origin of the lepton flavor mixing seems to be quite
different from that of the quark sector. 
However, in this Letter, we will show that the lepton
flavor mixing can be obtained via diagonalization of the mass matrix
based on the broken permutation symmetry, exactly as in the quark sector.
In order to do that, we will first derive the quark mixing matrix
from the mass matrix {\it Ansatz} based on the broken permutation symmetry.
Then, we will extend it to the lepton sector.
We will, in particular, show that the ``bimaximal" mixing of the lepton flavor
can be obtained in such a scheme.

Recently a new general class of mass matrix {\it Ansatz}, that respects 
the quark mass hierarchy of the quark flavor mixing matrix, 
has been studied. That is a generalization
of various specific forms of mass matrix by successive breaking
of the maximal permutation symmetry. 
The resulting mass matrix in the hierarchical basis is of 
the form \cite{Xing}:
\begin{equation}
M_H = \left(
   \begin{array}{ccc}
     0    &   A   & 0 \\
     A    &   D   & B \\
     0    &   B   & C
   \end{array}
   \right).
\end{equation}
The matrix $M_H$ contains four independent parameters even in the case of real
parameters so that the genuine {\it calculability} is lost. 
In order to maintain the  calculability, 
one has to make additional {\it Ansatz} to provide
any relationship between two of the four independent parameters, 
as shown in Ref. \cite{kk}.
In this Letter, however, we do not make additional {\it Ansatz} to keep 
the calculability but rather introduce additional parameter 
besides three fermion masses. This additional parameter will be determined
from the best fit to the measured quark CKM mixing matrix and the observed
neutrino mass hierarchy.

The parameters $A,B,C$ and $D$ can be expressed in terms of the fermion
mass eigenvalues.
In view of the hierarchical pattern of the quark masses, it is natural
to expect that $A<D<<C$, and then one can take the mass eigenvalues to be
$-m_1, m_2$ and $m_3$. 
The trace and the determinant of $M_H$ should be given, respectively,
\begin{eqnarray} 
Tr(M_H) &=& -m_1+m_2+m_3,~~~~~{\rm and}~~~~~ Det(M_H)= -m_1m_2m_3. \nonumber\\
Tr(M_H^2) &=& m_1^2+m_2^2+m_3^2, \nonumber
\end{eqnarray}
{}From those relations, we obtain the following form of fermion mass 
matrix:
\begin{equation}
M= \left(
   \begin{array}{ccc}
     0    &     \sqrt{\frac{m_1 m_2 m_3}{m_3-\epsilon}} & 0 \\
     \sqrt{\frac{m_1 m_2 m_3}{m_3-\epsilon}} & m_2-m_1+\epsilon
                           & \omega(m_2-m_1+\epsilon) \\
     0 & \omega(m_2-m_1+\epsilon)  & m_3-\epsilon
   \end{array}
   \right) ,
\label{mm}
\end{equation}
in which  the analytic relation between two parameters
$\epsilon$ and $\omega$ is given by
\begin{equation}
\omega^2=
   \frac{\epsilon (m_3-m_2+m_1-\epsilon)(m_3-\epsilon)-\epsilon m_1m_2}
        {(m_3-\epsilon)(m_2-m_1+\epsilon)^2} .
\label{omega2}
\end{equation}
For given three fermion masses, $\omega^2$ is 
determined by fixing the parameter $\epsilon$.
Therefore, the mixing matrix can be expressed
in terms of three fermion mass eigenvalues and additional parameter $\epsilon$.
Moreover we have to restrict the parameter range, $0<\epsilon<(m_3-m_2)$,
so that the mass matrix is to be real symmetric.
Although there are four independent parameters in the mass matrix, Eq. (2),
it will be shown in Sections 2 and 3 that by varying
the additional parameter $\epsilon$ we can reproduce not only the hierarchical
quark CKM mixing matrix but also the ``bimaximal" lepton mixing matrix
from the same mass matrix {\it Ansatz} of Eq. (2).

The real symmetric mass matrix $M$ can be diagonalized by
an orthogonal matrix $U$ as follows:
\begin{equation}
UMU^{\dagger}=\mbox{diag}(-m_1,m_2,m_3) . 
\end{equation}
The analytic formula for the orthogonal matrix is obtained, 
after simple algebra,
\begin{equation}
U= \left(
   \begin{array}{ccc}
      \sqrt{\frac{m_2m_3}{Y_1}}            (m_3+m_1-\epsilon) &
     -\sqrt{\frac{m_1(m_3-\epsilon)}{Y_1}} (m_3+m_1-\epsilon) &
      \sqrt{\frac{m_1(m_3-\epsilon)}{Y_1}}\omega(m_2-m_1+\epsilon) \\
      \sqrt{\frac{m_1m_3}{Y_2}}            (m_3-m_2-\epsilon) &
      \sqrt{\frac{m_2(m_3-\epsilon)}{Y_2}} (m_3-m_2-\epsilon) &
     -\sqrt{\frac{m_2(m_3-\epsilon)}{Y_2}}\omega(m_2-m_1+\epsilon) \\
      \sqrt{\frac{m_1m_2}{Y_3}}            \epsilon     &
      \sqrt{\frac{m_3(m_3-\epsilon)}{Y_3}} \epsilon     &
      \sqrt{\frac{m_3(m_3-\epsilon)}{Y_3}}\omega(m_2-m_1+\epsilon) \\
   \end{array}
   \right) ,
\end{equation}
where the normalization factors are
\begin{eqnarray}
Y_1 &=& (m_2m_3+m_1(m_3-\epsilon))(m_3+m_1-\epsilon)^2
        +m_1(m_3-\epsilon)\omega^2(m_2-m_1+\epsilon)^2    , 
\label{y1}   \\
Y_2 &=& (m_1m_3+m_2(m_3-\epsilon))(m_3-m_2-\epsilon)^2
        +m_2(m_3-\epsilon)\omega^2(m_2-m_1+\epsilon)^2    , 
\label{y2}  \\
Y_3 &=& (m_1m_2+m_3(m_3-\epsilon))\epsilon^2
        +m_3(m_3-\epsilon)\omega^2(m_2-m_1+\epsilon)^2  .
\label{y3}  
\end{eqnarray}

Using the above formulae, one can calculate the quark and 
lepton flavor mixing  matrices.
Since the flavor mixing matrix for the quark
sector is CKM matrix, we will determine the parameter $\epsilon$
from the best fit to the measured CKM matrix elements by using the running
quark masses at 1 GeV scale.
As shown in Section 2, the CP-violating phase $\delta$ will also be determined
from the best fit analysis.

In the lepton sector, the mixing matrix can be expressed in terms of
the lepton masses.
Thanks to the evidence for the neutrino oscillations and nonzero neutrino 
masses, one can phenomenologically construct the lepton mixing matrix 
in such a way as to be  consistent with the present neutrino experiments.
Preferring  the vacuum oscillation solutions for the solar and atmospheric
neutrino deficits, 
we may take into account the ``bimaximal" mixing scenario \cite{bimax}.
In Section 3, we will show how the  {\it near} ``bimaximal" mixing matrix can 
be achieved from our same mass matrix {\it Ansatz} by assuming
the hierarchy of neutrino masses to be $m_1\simeq m_2 <<m_3$
and taking parameter $\epsilon$ appropriately.

\section{Quark Flavor CKM Mixing Matrix}

Now let us consider the quark sector with the form of  mass matrix, Eq. (2).
We will take the up-type and down-type quark mass matrix as follows:
\begin{eqnarray}
M_u &=& M  ,    \\
M_d &=& PMP^{-1} ,
\end{eqnarray}
where the CP-violating phase $\delta$ is from 
$P=\mbox{diag}(\exp(i\delta),1,1)$,
and for $M_{u,d}$ the corresponding mass matrices are with eigenvalues
\begin{eqnarray}
(-m_u,m_c,m_t) && {\rm for~~ up~~type~~ quarks}, \nonumber \\ 
{\rm and} ~~~~~(-m_d,m_s,m_b) && {\rm for~~ down~~type~~ quarks}. \nonumber
\end{eqnarray}
Note that we have used the same parameter $\epsilon$
for up- and down-type quark mass matrices.

The CKM matrix can be obtained by
\begin{equation}
V_{\mbox{\tiny CKM}} = U_u P U_d^{\dagger} P^{-1} ,
\end{equation}
from the orthogonal matrices of $U_u,U_d$ of Eq. (4). 
Explicit formulae of the CKM matrix elements are as follows:
\begin{eqnarray}
V_{ud} \!\!\! &=&\!\!\! \{
      [\sqrt{m_c m_t m_s m_b}+
            e^{-i\delta}
     \sqrt{m_u(m_t-\epsilon)m_d(m_b-\epsilon)}]
     (m_t+m_u-\epsilon)(m_b+m_d-\epsilon)     \nonumber   \\
       & & 
     +e^{-i\delta}
     \sqrt{m_u(m_t-\epsilon)m_d(m_b-\epsilon)}
     \omega^u\omega^d(m_c-m_u+\epsilon)(m_s-m_d+\epsilon) \}
     /\sqrt{Y_1^u Y_1^d} ,       \\
V_{us} \!\!\! &=& \!\!\! \{
     [e^{i\delta}
     \sqrt{m_c m_t m_d m_b}
    -\sqrt{m_u(m_t-\epsilon)m_s(m_b-\epsilon)}]
     (m_t+m_u-\epsilon)(m_b-m_s-\epsilon)  \nonumber  \\
       & &
    -\sqrt{m_u(m_t-\epsilon)m_s(m_b-\epsilon)}
     \omega^u\omega^d(m_c-m_u+\epsilon)(m_s-m_d+\epsilon) \}
     /\sqrt{Y_1^u Y_2^d}  ,     \\
V_{ub} \!\!\! &=&\!\!\!  \{
     [e^{i\delta}
     \sqrt{m_c m_t m_d m_s}
    -\sqrt{m_u(m_t-\epsilon)m_b(m_b-\epsilon)}]
     (m_t+m_u-\epsilon)\epsilon            \nonumber \\
      & &
    +\sqrt{m_u(m_t-\epsilon)m_b(m_b-\epsilon)}
     \omega^u\omega^d(m_c-m_u+\epsilon)(m_s-m_d+\epsilon) \}
     /\sqrt{Y_1^u Y_3^d}  ,     \\  [.25in]
V_{cd} \!\!\! &=&\!\!\!  \{
      [\sqrt{m_u m_t m_s m_b}-
            e^{-i\delta}
     \sqrt{m_c(m_t-\epsilon)m_d(m_b-\epsilon)}]
     (m_t-m_c-\epsilon)(m_b+m_d-\epsilon)     \nonumber   \\
       & &
     -e^{-i\delta}
     \sqrt{m_c(m_t-\epsilon)m_d(m_b-\epsilon)}
     \omega^u\omega^d(m_c-m_u+\epsilon)(m_s-m_d+\epsilon) \}
     /\sqrt{Y_2^u Y_1^d}       ,      \\
V_{cs} \!\!\! &=&\!\!\!  \{
     [e^{i\delta}
     \sqrt{m_u m_t m_d m_b}
    +\sqrt{m_c(m_t-\epsilon)m_s(m_b-\epsilon)}]
     (m_t-m_c-\epsilon)(m_b-m_s-\epsilon)  \nonumber  \\
       & &
    +\sqrt{m_c(m_t-\epsilon)m_s(m_b-\epsilon)}
     \omega^u\omega^d(m_c-m_u+\epsilon)(m_s-m_d+\epsilon) \}
     /\sqrt{Y_2^u Y_2^d}  ,     \\
V_{cb} \!\!\! &=&\!\!\!  \{
     [e^{i\delta}
     \sqrt{m_u m_t m_d m_s}
    +\sqrt{m_c(m_t-\epsilon)m_b(m_b-\epsilon)}]
     (m_t-m_c-\epsilon)\epsilon            \nonumber \\
      & &
    -\sqrt{m_c(m_t-\epsilon)m_b(m_b-\epsilon)}
     \omega^u\omega^d(m_c-m_u+\epsilon)(m_s-m_d+\epsilon) \}
     /\sqrt{Y_2^u Y_3^d}    ,  \\   [.25in]
V_{td} \!\!\! &=&\!\!\!  \{
      [\sqrt{m_u m_c m_s m_b}-
            e^{-i\delta}
     \sqrt{m_t(m_t-\epsilon)m_d(m_b-\epsilon)}]
     \epsilon(m_b+m_d-\epsilon)     \nonumber   \\
       & &
     +e^{-i\delta}
     \sqrt{m_t(m_t-\epsilon)m_d(m_b-\epsilon)}
     \omega^u\omega^d(m_c-m_u+\epsilon)(m_s-m_d+\epsilon) \}
     /\sqrt{Y_3^u Y_1^d}   , \\
V_{ts} \!\!\! &=&\!\!\!  \{
     [e^{i\delta}
     \sqrt{m_u m_c m_d m_b}
    +\sqrt{m_t(m_t-\epsilon)m_s(m_b-\epsilon)}]
     \epsilon(m_b-m_s-\epsilon)  \nonumber  \\
       & &
    -\sqrt{m_t(m_t-\epsilon)m_s(m_b-\epsilon)}
     \omega^u\omega^d(m_c-m_u+\epsilon)(m_s-m_d+\epsilon) \}
     /\sqrt{Y_3^u Y_2^d}   ,  \\
V_{tb} \!\!\! &=&\!\!\!  \{
     [e^{i\delta}
     \sqrt{m_u m_c m_d m_s}
    +\sqrt{m_t(m_t-\epsilon)m_b(m_b-\epsilon)}]
     \epsilon^2            \nonumber \\
      & &
    +\sqrt{m_t(m_t-\epsilon)m_b(m_b-\epsilon)}
     \omega^u\omega^d(m_c-m_u+\epsilon)(m_s-m_d+\epsilon) \}
     /\sqrt{Y_3^u Y_3^d}  ,
\end{eqnarray}
where $Y_1^{u(d)},Y_2^{u(d)},Y_3^{u(d)},$ and $\omega^{u(d)}$ 
are given in Eqs.~(\ref{y1}),(\ref{y2}),(\ref{y3}),(\ref{omega2}), 
with fermion masses replaced by up-type
(down-type) quark masses.
Because the up-type quarks have mass hierarchy, $m_u \ll m_c \ll m_t$,
we can approximate the quantities 
$Y_1^u\simeq Y_2^u\simeq m_t^4(m_c/m_t)$ and 
$Y_3^u\simeq m_t^4(\epsilon/m_t)$.
The orthogonal matrix $U_u$, 
which diagonalizes the up-type quark mass matrix,
may be approximated
\begin{equation}
U_u \simeq
\left(
\begin{array}{ccc}
1   &   -\sqrt{\frac{m_u}{m_c}}   &   
                      \sqrt{\frac{m_u}{m_c}}\sqrt{\frac{\epsilon}{m_t}} \\
\sqrt{\frac{m_u}{m_c}}  &  1  &  -\sqrt{\frac{\epsilon}{m_t}}    \\
0   & \sqrt{\frac{\epsilon}{m_t}}   &   1
\end{array} 
\right)  .
\end{equation}
Similar approximations are applied to the down-type quarks as well
because of the observed similar mass hierarchy, $m_d \ll m_s \ll m_b$.     
Here we have presumed that $\epsilon \ll m_t, m_b$.

Due to the mass hierarchy in the quark sector,
we can express the CKM matrix elements in the leading approximation,
\begin{eqnarray}
V_{\mbox{\tiny us}} &\simeq& e^{i\delta}\sqrt{\frac{m_d}{m_s}}
                             -\sqrt{\frac{m_u}{m_c}} ,  \label{vus} \\
V_{\mbox{\tiny ub}} &\simeq& -\sqrt{\frac{m_u}{m_c}}
               (\sqrt{\frac{\epsilon}{m_b}} -\sqrt{\frac{\epsilon}{m_t}}) ,  \\
V_{\mbox{\tiny cb}} &\simeq& 
           \sqrt{\frac{\epsilon}{m_b}} -\sqrt{\frac{\epsilon}{m_t}} .
\end{eqnarray}
Note that our result for $|V_{\mbox{\tiny us}}|$ in Eq.~(\ref{vus}) is
the same as that of the Fritzsch model~\cite{wein}, and the ratio of
$|V_{\mbox{\tiny ub}}|$ to $|V_{\mbox{\tiny cb}}|$ is 
$|V_{\mbox{\tiny ub}}|/|V_{\mbox{\tiny cb}}|\simeq\sqrt{m_u/m_c}$.
The predictions for  $V_{\mbox{\tiny ub}}$ and $V_{\mbox{\tiny cb}}$ are
different from those of
the Fritzsch {\it Ansatz}.
The free parameter $\epsilon$ gives us flexibility to reproduce
measured values of $|V_{\mbox{\tiny ub}}|$ and $|V_{\mbox{\tiny cb}}|$
at the low energy scale, even if top quark mass is rather large.
We can also obtain the value of Jarlskog factor from the relation
\begin{equation}
J_{\mbox{\tiny CP}} \equiv
\mbox{Im} (V_{\mbox{\tiny ub}}  V_{\mbox{\tiny td}}
           V_{\mbox{\tiny ud}}^*V_{\mbox{\tiny tb}}^*) 
\simeq
\sqrt{\frac{m_u}{m_c}}\sqrt{\frac{m_d}{m_s}}
\frac{\epsilon}{m_b}\sin\delta    .
\end{equation}

For the numerical work,
we have used the running quark masses
at 1 GeV  scale~\cite{koide}:
\begin{eqnarray}
&& m_u=4.88\pm 0.57~{\rm MeV},~~~~~~ 
m_d=9.81\pm 0.65~{\rm MeV}, \nonumber \\
&& m_c=1.51\pm 0.04~{\rm GeV},~~~~~~~
m_s=195.4\pm 12.5~{\rm MeV}, \nonumber \\
{\rm and}~~ && m_t=475\pm 80~{\rm GeV},~~~~~~~~~~ 
m_b=7.2\pm 0.6~{\rm GeV}. \nonumber
\end{eqnarray}
The experimental values of CKM mixing matrix elements~\cite{meas,PDB} are
$$ 
|V_{us}|=0.2196\pm 0.0023,~~~~~
|V_{cb}|=0.0395\pm 0.0017,~~~~~
|V_{ub}/V_{cb}|=0.08\pm 0.02.
$$
We have made $\chi^2$ analysis to obtain two parameters, 
mass matrix parameter $\epsilon$ and the phase $\delta$, 
which induce the three observed experimental values of CKM matrix.
{}From numerical computation, 
we find that the best fit to the measured CKM 
matrix elements occurs at 
$$
\epsilon=14.2~{\rm MeV~~~~and}~~~~~
\delta=1.47~{\rm radians~~~~ with}~~~~~
\chi^2_{min}=1.36.
$$
For 90\% significance level, 
the two parameters are in the range of
$12.8\mbox{MeV}<\epsilon<15.5\mbox{MeV}$
and $1.42 < \delta < 1.52$.
The resulting CKM mixing matrix is then given, as in central values, 
\begin{equation}
V_{\mbox{\tiny CKM}} = \left(
   \begin{array}{ccc}
 0.9755-0.0124i & -0.0334+0.2172i & -0.0022+0.0003i   \\
 0.0335+0.2170i &  0.9747+0.0124i &  0.0395          \\
 0.0008-0.0083i & -0.0387         &  0.9992
   \end{array}
   \right)  .
\end{equation}
Note that the predicted ratio
$|V_{ub}/V_{cb}|$ prefers lower bound on the experimental measurement,
$\sim 0.06$.
The Jarlskog factor is estimated as 
\begin{equation}
J_{\mbox{\tiny CP}} = (0.18 \pm 0.02)\times 10^{-4}.
\end{equation} 
The moduli of the quark mixing matrix elements and the quantity
$J_{\mbox{\tiny CP}}$ do not depend on the parametrization chosen.
We can rewrite the CKM mixing matrix (26) with the standard parametrization
as used in the particle data book~\cite{PDB}:
\begin{equation}
V_{\mbox{\tiny CKM}} = \left(
   \begin{array}{ccc}
  0.9756         &  0.2197          & 0.0006-0.0022i   \\
 -0.2196         &  0.9748          & 0.0395          \\
  0.0081-0.0021i & -0.0387-0.0005i  & 0.9992
   \end{array}
   \right)  .
\end{equation}
Within the $1~ \sigma$ range, the mixing angles and the phase are 
corresponding to 
\begin{eqnarray}
\theta_{12}=(12.7 &\pm& 0.2)^{\circ},~~~
\theta_{23}=(2.3 \pm 0.2)^{\circ},~~~
\theta_{13}=(0.13 \pm 0.01)^{\circ}~~~{\rm and}~~~
\delta_{13}=(75 \pm 3)^{\circ}, \nonumber \\
\Longleftrightarrow~~~  
V_{us} &\simeq& \sin \theta_{12} =0.2197 \pm  0.0030, ~~~
V_{cb} \simeq \sin \theta_{23} = 0.0395 \pm 0.0020 ~~~{\rm and} \nonumber \\
V_{ub} &=& \sin \theta_{13} \exp(-i\delta_{13}) 
       = (0.0006 \pm 0.0002)-(0.0022 \pm 0.0002)i . \nonumber
\end{eqnarray}
We often use the unitary triangle for the study of CP violation.
Our model predicts that one of the angles of the unitary triangle is
$$
\beta = \mbox{arg}(-V_{cd}V_{cb}^*/V_{td}V_{tb}^*) 
   = (15 \pm 2)^{\circ},~~~ 
{\rm and}~~~\sin 2\beta = 0.50 \pm 0.06, 
$$
which is consistent with
the CDF results~\cite{CDF}, 
$\sin 2\beta=0.79 \footnotesize{
\begin{array}{c} +0.41  \\ -0.44 \end{array} } \normalsize $ .

\section{Lepton Mass Matrix and Neutrino Oscillation}

Now let us consider the lepton sector with the same mass matrix {\it Ansatz}
of Eq.~(\ref{mm}). We take the CP phase $\delta=0$, for a while,
so that CP is conserved in the lepton sector.
At present, the elements of the lepton flavor mixing matrix are determined
from the neutrino oscillation experiments.
Recent atmospheric neutrino experiments 
from Super-Kamiokande~\cite{superk1}
show evidence for neutrino oscillation 
and hence for a nonzero neutrino mass.
The results indicate the maximal mixing between the muon neutrino and tau 
neutrino states with a mass squared difference 
$$
\Delta m_{32}^2 \sim 2\times 10^{-3}~{\rm eV}^2.
$$
On the other hand, the solar neutrino deficit \cite{solar1,solar2} 
may be explained through matter enhanced neutrino oscillation 
[{\it i.e.}, the Mikheyev-Smirnov-Wolfenstein (MSW) solution \cite{msw}] if 
\begin{eqnarray}
\delta m^2_{solar} &\simeq& 6\times 10^{-6}~\mbox{eV}^2~~{\rm and}~~
\sin^2 2\theta_{solar}\simeq 7\times 10^{-3}~~
{\rm (small~~ angle~~ case)}, \nonumber\\
{\rm or}~~~ \delta m^2_{solar} &\simeq& 9\times 10^{-6}~\mbox{eV}^2~~{\rm and}
~~\sin^2 2\theta_{solar}\simeq 0.6~~{\rm (large~~ angle~~ case)}, \nonumber
\end{eqnarray}
and through the long-distance vacuum neutrino oscillation called
``just-so" oscillation \cite{justso} if 
$$
\delta m^2_{solar} \simeq 10^{-10}
~\mbox{eV}^2~~~~{\rm and}~~~~ \sin^2 2\theta_{solar} \simeq 1.0.
$$
However, the recent data on the electron neutrino spectrum reported by 
Super-Kamiokande \cite{solar2} seem to favor the ``just-so"
vacuum oscillation, 
even though the small angle MSW oscillation and the maximal mixing 
between the atmospheric $\nu_{\mu}$ and $\nu_{\tau}$ have been 
taken as a natural solution for the neutrino problems \cite{yana}.
Then, these results from neutrino experiments  imply
that three flavor neutrinos are oscillating along {\it bimaximal} mixing
pattern with the observed mass hierarchy, 
$\Delta m_{21}^2\ll \Delta m_{32}^2$ or $m_1\simeq m_2 \ll m_3$. 
Bimaximal neutrino mixing has been studied in the recent literature
\cite{bimax}.

Now we show how the {\it nearly} ``bimaximal" mixing in neutrino oscillation 
can be achieved from our mass matrix {\it Ansatz} of Eq. (2). 
The flavor (or weak) neutrino states are 
superposition of mass states and we may write
\begin{equation}
|\nu_{\alpha}>=\sum^3_{i=1} V^l_{\alpha i} |\nu_i> ,
\end{equation}
where $\alpha=e,\mu,\tau$, and index $i$ represents 
mass eigenstate of neutrino.
Let us denote the unitary matrices,
which make the neutrino mass matrix
$M_{\nu}$ and charged lepton mass matrix $M_l$ diagonal,
as $U_{\nu}$ and $U_l$,
\begin{eqnarray}
U_{\nu} M_{\nu} U_{\nu}^{\dagger} &=& \mbox{diag}(-m_1,m_2,m_3),\nonumber \\
U_l M_l U_l^{\dagger} &=& \mbox{diag}(-m_e,m_{\mu},m_{\tau}), \nonumber
\end{eqnarray}
where $m_1,m_2,$ and $m_3$ are neutrino masses hereafter, and
$M_{\nu}$ and $M_l$ have the matrix form of Eq.~(\ref{mm}).

The mixing matrix $V^l_{\mbox{\tiny CKM}}$ in neutrino oscillations 
is related to $U_{\nu}$ and $U_l$ as follows:
\begin{equation}
V^l_{\mbox{\tiny CKM}}=U_l^* U_{\nu}^T ,
\label{vlepton}
\end{equation}
where $T$ means the transpose.
Since the charged lepton family has mass hierarchy 
$m_e\ll m_{\mu}\ll m_{\tau}$, the approximate form of the orthogonal matrix
$U_l$ can be obtained, like in the quark case,
\begin{equation}
U_l \simeq
\left(
\begin{array}{ccc}
1   &   -\sqrt{\frac{m_e}{m_{\mu}}}   & 0 \\
\sqrt{\frac{m_e}{m_{\mu}}}  &  1  &  -\sqrt{\frac{\epsilon^l}{m_{\tau}}}  \\
0   & \sqrt{\frac{\epsilon^l}{m_{\tau}}}   &   1
\end{array}
\right)  ,
\end{equation}
where we assumed that $\epsilon^l \ll m_{\tau}$.
Now by choosing the parameter $\epsilon^l\simeq m_3/2$,
we take the neutrino mass matrix {\it Ansatz} such that
its masses satisfy the observed hierarchy of $m_1\simeq m_2 \ll m_3$.
And the neutrino mass matrix has form of Eq. (2) with
$\epsilon^l \simeq m_3/2$,
\begin{equation}
M_{\nu} \simeq \left(
   \begin{array}{ccc}
     0    &   \sqrt{2m_1 m_2}   & 0 \\
     \sqrt{2m_1 m_2}    &   m_3/2   & m_3/2  \\
     0    &   m_3/2    & m_3/2
   \end{array}
   \right) .
\end{equation} 
In this case we can also write $U_{\nu}$ approximately as follows,
\begin{equation}
U_{\nu} \simeq
\left(
\begin{array}{ccc}
\sqrt{\frac{m_2}{m_1+m_2}}   &   
-\sqrt{\frac{m_1}{m_1+m_2}}\sqrt{1-\frac{\epsilon^l}{m_3}}   & 
 \sqrt{\frac{m_1}{m_1+m_2}}\sqrt{\frac{\epsilon^l}{m_3}} \\
\sqrt{\frac{m_1}{m_1+m_2}}   &
\sqrt{\frac{m_2}{m_1+m_2}}\sqrt{1-\frac{\epsilon^l}{m_3}}   &
-\sqrt{\frac{m_2}{m_1+m_2}}\sqrt{\frac{\epsilon^l}{m_3}} \\
0   & \sqrt{\frac{\epsilon^l}{m_3}}   &  
\sqrt{1-\frac{\epsilon^l}{m_3}}
\end{array}
\right) .
\end{equation}
Then, using Eq.~(\ref{vlepton}), 
we can obtain expressions for the mixing matrix elements such as
\begin{eqnarray}
V^l_{e1} &\simeq& \sqrt{\frac{m_2}{m_1+m_2}}
         +\sqrt{\frac{m_e}{m_{\mu}}}
          \sqrt{\frac{m_1}{m_1+m_2}}\sqrt{1-\frac{\epsilon^l}{m_3}}  , \\
V^l_{e2} &\simeq& \sqrt{\frac{m_1}{m_1+m_2}}
         -\sqrt{\frac{m_e}{m_{\mu}}}
          \sqrt{\frac{m_2}{m_1+m_2}}\sqrt{1-\frac{\epsilon^l}{m_3}}  , \\
V^l_{e3} &\simeq& 
           -\sqrt{\frac{m_e}{m_{\mu}}}\sqrt{\frac{\epsilon^l}{m_3}}  ,
                       \\  [.1in]
V^l_{\mu 1} &\simeq& 
           -\sqrt{\frac{m_1}{m_1+m_2}}\sqrt{1-\frac{\epsilon^l}{m_3}} 
       +\sqrt{\frac{m_e}{m_{\mu}}} \sqrt{\frac{m_2}{m_1+m_2}} , \\
V^l_{\mu 2} &\simeq& 
           \sqrt{\frac{m_2}{m_1+m_2}}\sqrt{1-\frac{\epsilon^l}{m_3}} 
         + \sqrt{\frac{m_e}{m_{\mu}}} \sqrt{\frac{m_1}{m_1+m_2}} , \\
V^l_{\mu 3} &\simeq& \sqrt{\frac{\epsilon^l}{m_3}} , \\   [.1in]
V^l_{\tau 1} &\simeq& 
            \sqrt{\frac{m_1}{m_1+m_2}}\sqrt{\frac{\epsilon^l}{m_3}}, \\
V^l_{\tau 2} &\simeq& 
         -\sqrt{\frac{m_2}{m_1+m_2}}\sqrt{\frac{\epsilon^l}{m_3}}, \\
V^l_{\tau 3} &\simeq& \sqrt{1-\frac{\epsilon^l}{m_3}}  .
\end{eqnarray} 
As one can easily see, the ``{\it bimaximal}" mixing is 
nearly achieved when we set $\epsilon^l\simeq m_3/2$ 
in the lepton sector. 
Explicit lepton flavor mixing matrix $V^l_{\mbox{\tiny CKM}}$ 
may be written as
\begin{equation}
V^l_{\mbox{\tiny CKM}}\simeq
\left(
\begin{array}{ccc}
\sqrt{\frac{1}{2}}+\frac{1}{2}\sqrt{\frac{m_e}{m_{\mu}}} &
\sqrt{\frac{1}{2}}-\frac{1}{2}\sqrt{\frac{m_e}{m_{\mu}}} &
-\sqrt{\frac{1}{2}}\sqrt{\frac{m_e}{m_{\mu}}}   \\
-\frac{1}{2}+\frac{1}{2}\sqrt{\frac{m_e}{m_{\mu}}} &
 \frac{1}{2}+\frac{1}{2}\sqrt{\frac{m_e}{m_{\mu}}} &
 \sqrt{\frac{1}{2}}  \\
\frac{1}{2}  &
-\frac{1}{2}  &  \sqrt{\frac{1}{2}}  
\end{array}  
\right).
\end{equation}
Notice that the value of $V^l_{e3}$ is not exactly zero but small,
\begin{equation}
|V^l_{e3}| \simeq \frac{1}{\sqrt{2}}\sqrt{\frac{m_e}{m_{\mu}}} \simeq 0.05,
\end{equation}
which is consistent
with the bound obtained from CHOOZ experiment \cite{chooz} 
$|V^l_{e3}|\leq 0.22$ if $(m^2_3-m^2_1) > 10^{-3}~~\mbox{eV}^2$.
Although the exact ``bimaximal" neutrino mixing matrix predicts zero 
for the $V^l_{e3}$ element, 
a nonvanishing small $V^l_{e3}$ element is not completely excluded.
Since the $\nu_{\mu} \rightarrow \nu_{e}$ appearance channel is sensitive
to the product $|V_{\mu 3}^l V_{e3}^l|^2$ and  
$\nu_{\mu} \rightarrow \nu_{\tau}$ 
appearance channel is sensitive only to $|V^l_{\mu 3}|^2$, we can determine
the element $V^l_{e3}$ by combining the regions to be probed in both channels.
We expect that this will be  performed  in the future experiments 
such as K2K and MINOS.

Finally after including the CP-violating leptonic phase $\delta^l$, 
we can also calculate the theoretical upper bound of rephasing-invariant 
quantity analogous  to the Jarlskog invariant in the quark sector,
\begin{equation}
J^l_{\mbox{\tiny CP}} \simeq
\frac{\sqrt{m_1 m_2}}{m_1+m_2}
\frac{\epsilon^l}{m_3}
\sqrt{1-\frac{\epsilon^l}{m_3}}
\sqrt{\frac{m_e}{m_{\mu}}}\sin\delta^l
\leq 0.012.
\end{equation}
Compared to the quark sector result, $J_{\mbox{\tiny CP}}$ of Eq. (27), 
we find that $J^l_{\mbox{\tiny CP}}$ can be surprisingly large,
and the maximum amount of CP violation in the generation of the 
lepton number violation of the universe may be indeed large \cite{Sarkar}.
\\

\noindent
{\Large \bf Acknowledgments}
\\

\noindent
We thank G. Cvetic and S.K. Kang for careful reading of the manuscript 
and their valuable comments.
C.S.K. wishes to acknowledge the financial
support of 1997-sughak program of 
Korean Research Foundation, Project No. 1997-011-D00015.
The work of J.D.K. was supported in part by a postdoctoral grant
from the Natural Science Research Institute, Yonsei University in 1999.


\newpage

\begin{thebibliography}{99}

\bibitem{wein} H. Fritzsch, Phys. Lett. 
    {\bf B73}, 317 (1978); Nucl. Phys. {\bf B155}, 189 (1979).

\bibitem{cal} A.C. Rothman and K.S. Kang, 
  Phys. Rev. Lett. {\bf 43}, 1548 (1979).

\bibitem{test} 
K.S. Kang and S. Hadjitheodoridis, Phys. Lett. {\bf B193}, 504 (1987);
H. Harari and Y. Nir, Phys. Lett. {\bf B195}, 586  (1987).

\bibitem{cuypers} F. Cuypers and C.S. Kim, Phys. Lett. {\bf B254}, 462 (1991);
and the references therein.

\bibitem{km} M. Kobayashi and T. Maskawa, 
 Prog. Theor. Phys. {\bf 49}, 625 (1973).

\bibitem{wolf} L. Wolfenstein, Phys. Rev. Lett. {\bf 51}, 1945 (1983).

\bibitem{solar1} B.T. Cleveland {\it et al.}, Nucl. Phys. B (Proc. Suppl.) 
{\bf 38}, 47 (1995); 
Kamiokande Collaboration: Y. Fukuda {\it et al.}, 
 Phys. Rev. Lett. {\bf 77}, 1683 (1996); 
GALLEX Collaboration: W. Hampel {\it et al.}, 
 Phys. Lett. {\bf B388}, 384 (1996); 
SAGE Collaboration: J.N. Abdurashitov {\it et al.}, 
 Phys. Rev. Lett. {\bf 77}, 4708 (1996).

\bibitem{solar2} Super-Kamiokande Collaboration: 
 talk by Y. Suzuki at {\it Neutrino-98}, Takayama, Japan (June, 1998).

\bibitem{superk1} Super-Kamiokande Collaboration: Y. Fukuda, {\it et al.}, 
 Phys. Rev. Lett. {\bf 81}, 1562 (1998).

\bibitem{bimax} V. Barger, S. Pakvasa, T.J. Weiler and K. Whisnant, 
 Phys. Lett. {\bf B437}, 107 (1998); 
H. Georgi and S.L. Glashow, hep-ph/9808293;
S.K. Kang and C.S. Kim, Phys. Rev. {\bf D59}, 091302 (1999);
M. Jezabek and Y. Sumino,  Phys. Lett. {\bf B457}, 139 (1999);
Y. Wu, hep-ph/9905222; C.H. Albright and S.M. Barr, hep-ph/9906292;
and the references therein. 

\bibitem{Xing}
H. Fritzsch and Z. Xing, Phys. Lett. {\bf B353}, 114 (1995);
P. Kauss and S. Meshkov, Phys. Rev. {\bf D42}, 1863 (1990).

\bibitem{kk} K.S. Kang and S.K. Kang, Phys. Rev. {\bf D56}, 1511 (1997);
K.S. Kang, S.K. Kang, C.S. Kim and S.M. Kim, hep-ph/9808419.

\bibitem{msw} L. Wolfenstein, Phys. Rev. {\bf D17}, 2369 (1978); 
S. P. Mikheyev and A. Smirnov, Yad. Fiz. {\bf 42}, 1441 (1985); 
Nuovo Cimento {\bf 9C}, 17 (1986).

\bibitem{justso} 
V. Barger, R.J. Phillips and K. Whisnant, Phys. Rev. {\bf D24}, 538 (1981);
S.L. Glashow and L.M. Krauss, Phys. Lett. {\bf B190}, 199 (1987);
A.S. Joshipura and  M. Nowakowski, Phys. Rev. {\bf D51}, 2421 (1995). 

\bibitem{yana} T. Yanagida, talk at {\it Neutrino-98}, Japan (June, 1998);
 P. Ramond, talk at {\it Neutrino-98}, Japan (June, 1998).

\bibitem{koide}
H. Fusaoka and Y. Koide, Phys. Rev. {\bf D57}, 3986 (1998).

\bibitem{meas}
OPAL Collaboration: K. Ackerstaff {\it et al.}, 
 Phys. Lett. {\bf B395}, 128 (1997);
ALEPH Collaboration: D. Buskulic {\it et al.}, 
 Phys. Lett. {\bf B395}, 373 (1997).

\bibitem{PDB}
Particle Data Group, Eur. Phys. J. {\bf C3}, 103 (1998).

\bibitem{CDF}
CDF Collaboration: CDF/PUB/BOTTOM/CDF/4855.

\bibitem{chooz} 
The CHOOZ Collaboration: M. Apollonio {\it et al.}, 
 Phys. Lett. {\bf B420}, 397 (1998);
S.M. Bilenky and C. Giunti, Phys. Lett. {\bf B444}, 379 (1998).

\bibitem{Sarkar} Y. Liu and U. Sarkar, hep-ph/9906307.

\end{thebibliography}
\end{document}